\documentstyle[aps,floats,psfig,twocolumn]{revtex}

\begin{document}
\tighten
\draft
\title{Anisotropy of the Energy Gap in the Insulating Phase of the 
$\mbox{\boldmath$U-t-t'$}$ Hubbard Model}

\author{Ph. Brune \and A.~P. Kampf}
\address{Institut f\"ur Physik, Theoretische Physik III, 
Elektronische Korrelationen und Magnetismus,\\
Universit\"at Augsburg, 86135 Augsburg, Germany}
\address{~
\parbox{14cm}{\rm 
\medskip
We apply a diagrammatic expansion method around the atomic limit 
($U >> t$) for the $U$-$t$-$t'$ Hubbard model at half filling and finite 
temperature by means of a continued fraction representation of the 
one-particle Green's function. From the analysis of the spectral function 
$A(\vec{k},\omega)$ we find an energy dispersion relation with a 
$(\cos k_x-\cos k_y)^2$ modulation of the energy gap in the insulating phase. 
This anisotropy is compared with experimental ARPES results on insulating 
cuprates.
\vskip0.05cm\medskip 
PACS numbers: 71.10.Fd, 71.27.+a, 71.30.+h
}}	

\maketitle

During the last decade of research work on high temperature superconductors 
angular resolved photoemission spectroscopy (ARPES) has played an important 
role in elucidating their electronic properties \cite{ShenDessau}. An 
understanding of the single particle properties in the normal state is a 
prerequisite for a theory of a mechanism for superconductivity as well as for 
transport properties. Over the years ARPES data 
have continuously provided surprises and new insights and thereby served as a 
guidance to theoretical developments. The observation of the 
$d_{x^2-y^2}$-shape of the energy gap, pseudogap structures in the metallic 
phase \cite{Marshall,Ding}, strong anisotropies in the quasiparticle (qp)
peak lineshapes and a possible partial destruction of the Fermi surface 
\cite{Norman}, or the unusual frequency and temperature dependence of the 
qp peakwidth \cite{Valla} are examples for intriguing information obtained 
from ARPES experiments. 

For the normal state an important goal is the description for the evolution 
with doping from the antiferromagnetic (AF) and insulating parent compounds to 
the overdoped superconductors. This demands control over the spectral features 
of the Mott-Hubbard insulating state as a starting point. Yet, it proved to be 
difficult to reproduce ARPES data for the single hole dispersion in the 
insulating cuprate Sr$_2$CuO$_2$Cl$_2$ \cite{Wells95} within $t$-$J$ or 
Hubbard models. In particular, for momenta along the Brillouin zone (BZ) axis 
next-nearest neighbor (nnn) or even longer range hopping amplitudes had to be 
introduced to achieve a reasonable comparison to the measured spectra 
\cite{Duffy95,Nazarenko,Eder97,LemaLeung}. Even more striking are recent 
results on AF Ca$_2$CuO$_2$Cl$_2$ that demonstrated an 
anisotropy of the insulating energy gap which was claimed to follow closely a 
$d_{x^2-y^2}$-wave modulation along a remnant Fermi surface with a modulation 
amplitude of $\sim$ 300meV comparable in magnitude to the AF exchange 
interaction \cite{Ronning98,Ronning00}.
It has been pointed out that a $d_{x^2-y^2}$ gap modulation might follow 
naturally in the context of a projected SO(5) theory \cite{Hanke99} unifying 
antiferromagnetism and d-wave superconductivity by a symmetry principle 
\cite{Zhang}.

In this paper we show that a $\gamma^2_d(\vec{k})=(\cos k_x-\cos k_y)^2$ 
modulation of the AF energy gap is realized in the half-filled Hubbard model 
on a square lattice with nearest (nn) and nnn hopping amplitudes. This result 
is obtained in an analytic strong coupling expansion around the atomic limit 
following a recently proposed strategy with a mapping to a Jacobi continued 
fraction representation for the propagator \cite{Pairault98}. Given that the 
Mott-Hubbard insulator is the appropriate starting point for studying the hole 
doping evolution in cuprates this intrinsic anisotropy may bear an important 
preformed structure for the doped metallic phase.  

The single particle Matsubara Green's function for the Hubbard model is 
represented in terms of Grassmann fields $\gamma^*$ and $\gamma$ in the 
Feynman path integral representation by
\begin{eqnarray}
\nonumber 
G_{ij}(\tau\sigma|\tau'\sigma')
       =-\langle T^{\phantom{\dagger}}_{\tau}c^{\phantom{\dagger}}_{i\sigma}
(\tau)\, &&c^\dagger_{j\sigma'}(\tau')\rangle \\
={1\over Z}\int[d\gamma^* \, d\gamma]\,
         \gamma^{\phantom{*}}_{i\sigma\tau}\,\gamma^*_{j\sigma'\tau'}&&
         \,\exp(-S[\gamma^*,\gamma])
\label{G}
\end{eqnarray}
with the partition function $Z$ and the action $S=S_{kin}+S_{atom}$ where
\begin{eqnarray}
\label{S_kin}
S_{kin}[\gamma^*,\gamma]=-\int^{\beta}_0 d\tau\,
\sum_{i,j,\sigma}
t_{ij}(\gamma^{*}_{i\sigma\tau}\gamma^{\phantom{*}}_{j\sigma\tau}+h.c.) ,
\end{eqnarray}
\begin{eqnarray}
\nonumber
S_{atom}[\gamma^*,\gamma]=\int^{\beta}_0 d\tau\,
\biggl[\,&&\sum_{i,\sigma}\gamma^{*}_{i\sigma\tau}(\frac{\partial}
{\partial\tau}-\mu)
                \,\gamma^{\phantom{*}}_{i\sigma\tau}\phantom{\biggr]}\\
\phantom{\biggl[}
+U&&\sum_i \gamma^{*}_{i\uparrow\tau}\gamma^{*}_{i\downarrow\tau}
           \gamma^{\phantom{*}}_{i\downarrow\tau}
           \gamma^{\phantom{*}}_{i\uparrow\tau}\,\biggr] \, .
\end{eqnarray}
Here, $c^{\dagger}_{i\sigma}$ creates an electron at site $i$ with spin 
$\sigma$, and we restrict the hopping amplitudes $t_{ij}$ to nn ($t$) and nnn 
($t'$) sites. $U$ is the on-site Coulomb repulsion, 
$\mu$ the chemical potential, and $\beta=1/T$ the inverse temperature.

\begin{figure}
\vspace{10mm}
\centerline{\psfig{file=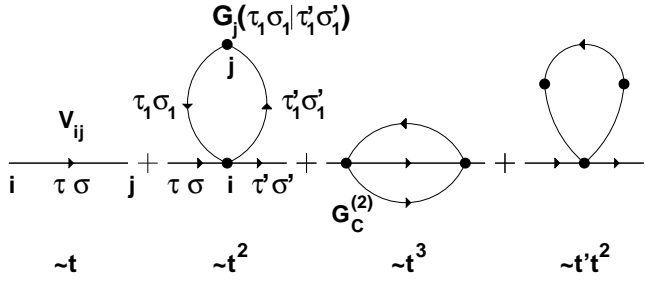,width=85mm,silent=}}
\vspace{5mm}
\caption{Irreducible diagrams contributing to $G(\vec{k},i\omega_n)$
of the $U$-$t$-$t'$ Hubbard model up to the order $(t/U)^3$.
The vertices ($\bullet$) represent local n-particle cumulants
$G^{(n)}_C$. Each of the lines represents a 
hopping process with amplitude either $t$ or $t'$.}
\label{fig:diagram33}
\end{figure}

$G_{ij}(\tau\sigma|\tau'\sigma')$
is evaluated diagrammatically in terms of a
cumulant expansion around the atomic limit, i.e. an expansion in powers of 
$t/U$ \cite{Metzner91}. As early on suggested by Sarker \cite{Sarker88} it 
is convenient to introduce auxiliary Grassmann fields 
$\{\psi^*_{i\sigma\tau},\psi^{\phantom{*}}_{i\sigma\tau}\}$ by performing a 
Hubbard-\-Stra\-to\-no\-vich transformation with respect to the kinetic energy term 
Eq. \ref{S_kin}. From the Green's function for the auxiliary 
fields ${\mathcal V}_{ij}(\tau\sigma|\tau'\sigma')=-\langle T_{\tau}
\psi^{\phantom{*}}_{i\sigma\tau}\psi^*_{j\sigma'\tau'}\rangle$ the propagator 
$G_{ij}(\tau\sigma|\tau'\sigma')$ in Eq. \ref{G} is obtained via the relation
$G_{ij}(\tau\sigma|\tau'\sigma')=(\Gamma^{-1}-V)^{-1}_{ij}(\tau\sigma|\tau'
\sigma')$ \cite{Pairault98}. Here $\Gamma$ denotes the
self energy and $V$ is the non-interacting Green's function for the auxiliary 
fields, i.e. ${\mathcal V}=V+V\Gamma{\mathcal V}$.

As is known from previous work on strong-coupling expansions for the Hubbard 
model \cite{Metzner91,BeckerFulde} the Green's function calculated this way 
does not have the correct analytic properties, because higher order cumulants
contain also higher order poles.
This problem was circumvented  in Ref. \cite{Pairault98} by mapping the 
Green's function as calculated by the diagrammatic expansion to a Green's 
function $G_J$ in a finite Jacobi continued fraction representation 
\cite{WallsBook}
\begin{equation}
G_J(z) = \displaystyle{a_0 \over z+b_1-\displaystyle{a_1 \over
                                 z+b_2-\displaystyle{a_2 \over z+\dots
                                      -\displaystyle{a_{N-1} \over z+b_N}}}}
\label{gj}
\end{equation}
such that $G_J$ has the same series expansion as $G$ to the same order in 
$t/U$.

Specifically, we have calculated the Fourier transformed Green's function 
$G(\vec{k},i\omega_n)$ at half filling $\mu=\frac{U}{2}$ up to order 
$(t/U)^4$. While our independently obtained results agree with 
Ref. \cite{Pairault98} for nn hopping only, in the case of a finite nnn 
hopping $t'$ an additional 
$tt'^2/U^3$ diagram contributes to $G(\vec{k},i\omega_n)$ as shown in Fig. 
\ref{fig:diagram33}. In this type of diagrams the vertices represent local 
n-particle cumulants $G^{(n)}_C$ (connected Green's functions). Each line 
between two vertices represents a hopping process between two sites. In the 
absence of a Wick theorem for the local Green's functions 
the higher order cumulants must be calculated separately. 
The resulting algebraic expressions become very involved for higher
order cumulants; for the performance of this diagrammatic expansion we have 
therefore developed a special purpose computer algebra code.

\begin{figure}
\vspace{10mm}
\centerline{\psfig{file=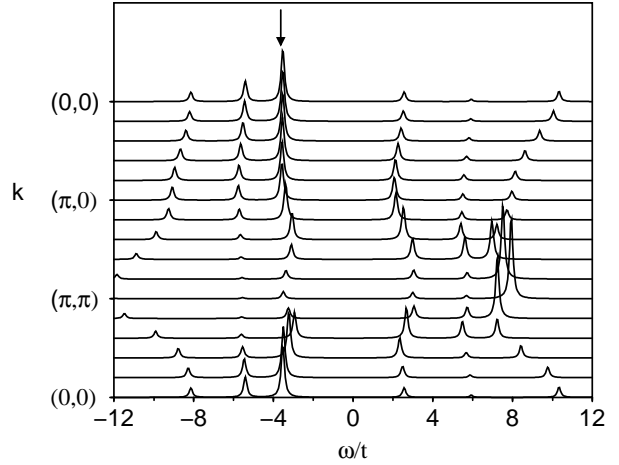,height=70mm,silent=}}
\vspace{5mm}
\caption{Spectral function $A(\vec{k},\omega)$ of the half-filled $U$-$t$-$t'$ 
Hubbard model as a function of $\omega$ along the path $\vec{k}=(0,0)$ 
$\rightarrow$ $(\pi,\pi)$ $\rightarrow$ $(\pi,0)$ $\rightarrow$ $(0,0)$ 
through the first Brillouin zone for $U=10t$, $T=0.2t$ and $t'=-0.45t$
as obtained from the strong coupling expansion to order $(t/U)^4$.}
\label{fig:A_k}
\end{figure}

As a result we obtain a Jacobi continued fraction expression for $G_J$ that has
eight fraction levels, i.e. the coefficients $a_N$ in Eq. \ref{gj} vanish for 
$N>8$. This termination for $G_J$ translates into eight simple poles. Each of 
the continued fraction coefficients $a_i$, $b_i$ $(i=1,\dots,8)$ is given by a 
fourth order polynomial in $t/U$ or $t'/U$, respectively, with coefficients 
that depend on $T$, $\mu$, $U$, and $\vec{k}$. The explicit result for $G_J$ 
is accessible electronically \cite{coefftable}. In Fig. \ref{fig:A_k} we show 
the corresponding spectral function 
\begin{equation}
A(\vec{k},\omega)=-\frac{1}{\pi}\,
                  {\mathrm Im}\,G(\vec{k},\omega+i0^+)
\end{equation}
for different $\vec{k}=(k_x,k_y)$ along the path 
$(0,0)\rightarrow(\pi,\pi)\rightarrow(\pi,0)\rightarrow(0,0)$ in the first 
BZ. On the scale of this plot only six out of eight poles of 
$G_J(\vec{k},\omega)$ carry significant and visible spectral weight.

\begin{figure}
\vspace{10mm}
\centerline{\psfig{file=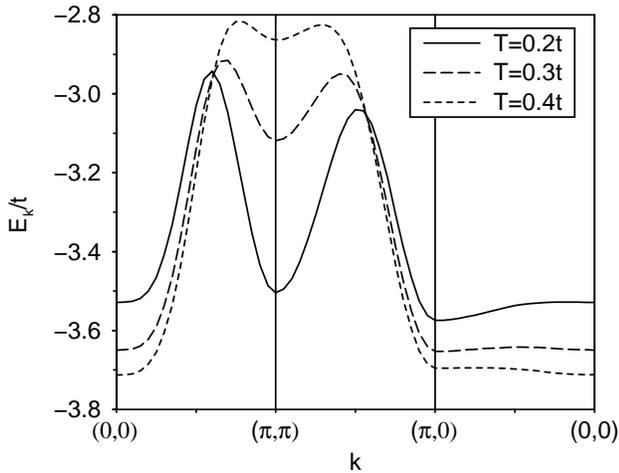,height=70mm,silent=}}
\vspace{5mm}
\caption{Temperature dependence of the dispersion $E(\vec{k})$ of the low 
energy peak (marked by an arrow in Fig. \ref{fig:A_k}) along a selected 
path in the first Brillouin zone for $U=10t$ and $t'=-0.45t$.}
\label{fig:Tdep}
\end{figure}

In analyzing the spectrum we map out the dispersion of the lowest energy hole 
excitation, i.e. the dispersion of the first peak below the gap (marked by an 
arrow in Fig. \ref{fig:A_k}). Its spectral weight is largest at the BZ
center and drops monotonically towards $(\pi,\pi)$. Fig. \ref{fig:Tdep} shows 
the peak dispersion for different temperatures. Upon cooling the system 
approaches the AF ordered ground state with a doubled unit 
cell and a reduced magnetic BZ (determined by $\cos k_x+\cos k_y\geq 0$). 
The dispersion along the zone diagonal approaches a perfectly symmetric shape 
with respect to the point $\vec{k}_d=(\frac{\pi}{2},\frac{\pi}{2})$ on the 
magnetic BZ boundary reflecting the growing AF spin correlations. Remarkably, 
along the BZ axis the dispersion remains very flat at all temperatures. In 
fact, the dispersion along the axis is obtained flatter than in previous 
numerical studies of Hubbard or $t$-$J$ models 
\cite{Duffy95,Eder97,Preuss97,Duffy97}. 
In units of the exchange coupling $J=4t^2/U$ the total bandwidth of the energy 
dispersion is $1.58$ for $T=0.2t$, which is roughly consistent with experiment 
and previous calculations.

\begin{figure}
\vspace{10mm}
\centerline{\psfig{file=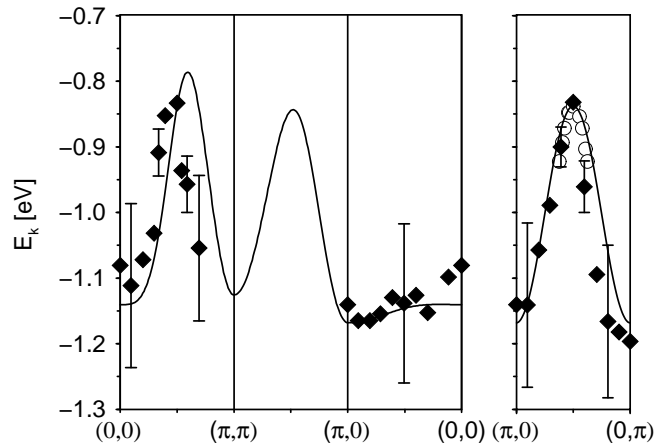,height=70mm,silent=}}
\vspace{5mm}
\caption{
A comparison with the ARPES dispersion for Sr$_2$CuO$_2$Cl$_2$ taken from 
Ref. [6] (black diamonds) and Ref. [12] (white circles). Error bars are shown 
at some selected experimental data points. The parameters in the theory were 
chosen as $U=10t$, $T=0.2t$, $t'=-0.45t$, and $t=0.605$eV.}
\label{fig:ExpFit}
\end{figure}

In Fig. \ref{fig:ExpFit} we compare our result for the single hole dispersion
with the ARPES data on Sr$_2$CuO$_2$Cl$_2$ from Ref. \cite{Wells95}. For
$U=10t$ the parameters $t$ and $t'$ were chosen to obtain a best fit to 
the data. 
The overall agreement, in particular along the BZ axis, is quite
satisfactory. Hopping amplitudes beyond nnn hopping are found unnecessary for 
reproducing the flat dispersion along the BZ axis.The deviation of the 
theoretical result from the experimental data points along the BZ diagonal, 
i.e. the lack of symmetry with respect to $\vec{k}_d$ -- as 
realized in the AF state -- is due to the finite temperature. In fact, the 
range of applicability of the $t/U$ expansion has a lower bound in temperature.
(Following the arguments of Ref. \cite{Pairault98} we estimate that our 
results are valid for $T>0.16t$.) When the magnetic correlation length is much 
larger than the hopping range as given by the order of the $t/U$ expansion, an 
accurate description of the single particle Green's function is no longer 
expected. The data were taken 100K above the N\'eel temperature $T_N=256$K of
Sr$_2$CuO$_2$Cl$_2$; at this temperature the magnetic
correlation length is already as large as 250{\AA} as measured by neutron 
scattering \cite{Greven}. Nevertheless, the flatness of the dispersion along 
the BZ axis in the $t/U$ expansion is a robust and temperature insensitive 
feature. 

\begin{figure}
\vspace{10mm}
\centerline{\psfig{file=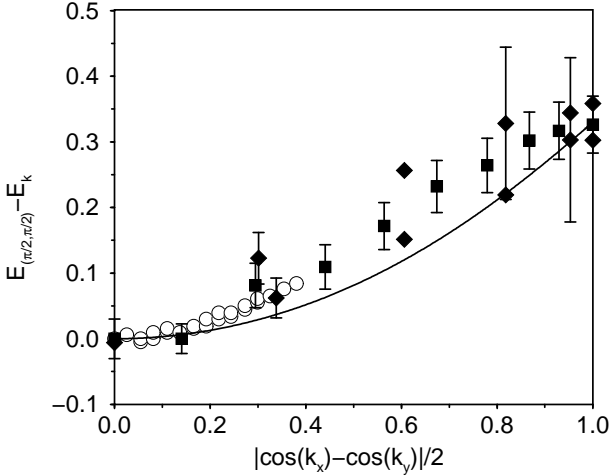,height=70mm,silent=}}
\vspace{5mm}
\caption{Same data as in Fig. \ref{fig:ExpFit} along the BZ path from 
$\vec{k}_d=(\pi/2,\pi/2)$ to $(\pi,0)$, but here $E(\vec{k}_d)-E(\vec{k})$ is
plotted as a function of $|\gamma_d(\vec{k})|/2$. In addition, ARPES
data on Ca$_2$CuO$_2$Cl$_2$ (from Ref. [11]) are shown (black squares).} 
\label{fig:ExpFit_dwave}
\end{figure}

In Fig. \ref{fig:ExpFit_dwave} the calculated energy dispersion 
$E(\vec{k}_d)-E(\vec{k})$ of the lowest energy peak and the experimental
data are plotted along the BZ path from $\vec{k}_d$ to 
$(\pi,0)$ which is parametrized in the form $|\gamma_d(\vec{k})|/2$. 
Originally, the data of Ref. \cite{Ronning98} (black squares) were
anticipated to imply a $d_{x^2-y^2}$-modulation of the energy gap which
in the pa\-ra\-me\-tri\-za\-tion of Fig. \ref{fig:ExpFit_dwave} would 
translate into a straight line. The perfectly linear relation between 
$E(\vec{k}_d)-E(\vec{k})$ and $|\gamma_d(\vec{k})|$ along the chosen
path would furthermore imply an unreasonable cusp-like feature of the 
dispersion on the magnetic BZ boundary. Indeed, the more recent data by
Ronning et al. \cite{Ronning00} (white circles in Fig.
\ref{fig:ExpFit_dwave}) resolve this problem and rather provide evidence 
for a quadratic dependence of the energy gap on $|\gamma_d(\vec{k})|$ in 
the vicinity of $\vec{k}_d$. Given the experimental error bars
of Ref. \cite{Wells95} included in Fig. \ref{fig:ExpFit_dwave} and in the
absence of an error estimate for the recent data by Ronning et al.
\cite{Ronning00} our results are clearly compatible with experiment.
With the same parameter set used for the fit of the ARPES dispersion in 
Fig. \ref{fig:ExpFit} the gap modulation amplitude $\sim$ 300meV as measured 
in Ca$_2$CuO$_2$Cl$_2$ \cite{Ronning98} is reproduced as well.

\begin{figure}
\vspace{10mm}
\centerline{\psfig{file=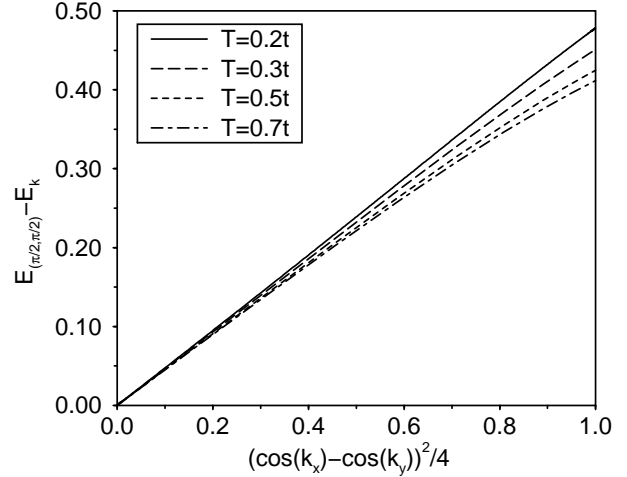,height=70mm,silent=}}
\vspace{5mm}
\caption{Dispersion along the path from $\vec{k}_d=(\pi/2,\pi/2)$ to 
$(\pi,0)$ relative to its value at $\vec{k}_d$ as a function of 
$(\cos k_x-\cos k_y)^2/4$ for different temperatures with $U=10t$ and 
$t'=-0.4t$.}
\label{fig:Tdep_mod}
\end{figure}

Fig. \ref{fig:Tdep_mod} shows the temperature dependence of the gap
modulation. In fact, when plotted versus $\gamma^2_d(\vec{k})$ it becomes
evident that the gap approaches a perfect $\gamma^2_d(\vec{k})$ momentum
dependence at low temperatures. For $t'=0$ the modulation 
vanishes in the expansion up to the order $(t/U)^4$ because the $\vec{k}$ 
dependence of all diagrams to this order arises in the form 
$\cos k_x+\cos k_y$ for $t'=0$ which vanishes on the BZ boundary
for $t'=0$. If diagrams of order $(t/U)^6$ or higher were taken into account,  
a small modulation is expected to appear also for $t'=0$.

In summary, we have found in a strong coupling expansion to order $(t/U)^4$ 
that the energy gap in the insulating phase of the half-filled $U$-$t$-$t'$ 
Hubbard model develops a $(\cos k_x-\cos k_y)^2$ modulation at low 
temperatures. 
Without the need to include hopping amplitudes beyond nnn an excellent fit
is achieved for the single hole dispersion in Sr$_2$CuO$_2$Cl$_2$.
With the same dispersion fit parameters also the calculated gap modulation 
amplitude compares well with the ARPES data. It is natural to expect that 
these spectral features in the insulator carry over to the metallic, doped 
case; the consequences and the connection to anisotropic pseudogap structures 
or even $d$-wave superconductivity remain yet to be understood. 

We thank D. Duffy and F. Ronning for discussions and S. Pairault for sharing 
his insight into the efficient calculation of high order diagrams. This work 
was supported by the Deutsche Forschungsgemeinschaft through SFB 484. 

\end{document}